\documentclass[a4paper,11pt]{article}
\pdfoutput=1 % if your are submitting a pdflatex (i.e. if you have
\usepackage{jheppub} % for details on the use of the package, please
\usepackage{graphicx}
\usepackage{epsfig}
\usepackage{rotating}
\usepackage{amssymb}
\usepackage{subfigure}
\usepackage{dsfont}
\usepackage{psfrag}
\usepackage{amsmath,euscript,array,mathrsfs}
\usepackage{axodraw}
\usepackage{bbold}
\usepackage{epsf}

\newcommand{\beq}{\begin{equation}}
\newcommand{\eeq}{\end{equation}}
\newcommand{\beqs}{\begin{eqnarray}}
\newcommand{\eeqs}{\end{eqnarray}}

\newcommand{\Tr}{{\rm Tr}}

\def\hbar{\hspace{0pt}\raisebox{1pt}{$-$} \hspace{-7pt} h}

\newcommand{\be}{\begin{equation}}
\newcommand{\ee}{\end{equation}}

\newcommand{\bea}{\begin{eqnarray}}
\newcommand{\eea}{\end{eqnarray}}

\def\lbldef#1#2{\expandafter\gdef\csname #1\endcsname {#2}}

\def\href#1#2{#2}

\newcommand{\ber}{\begin{eqnarray}}
\newcommand{\eer}{\end{eqnarray}}

\newcommand{\beqar}{\begin{eqnarray}}

\newcommand{\eeqar}{\end{eqnarray}}

\newcommand{\dsl}
  {\kern.06em\hbox{\raise.15ex\hbox{$/$}\kern-.56em\hbox{$\partial$}}}

\newcommand{\eeqarr}{\end{eqnarray}}
\newcommand{\ZZ}{{\rm \kern 0.275em Z \kern -0.92em Z}\;}

\def\CC{{\mathchoice
{\rm C\mkern-8mu\vrule height1.45ex depth-.05ex
width.05em\mkern9mu\kern-.05em}
{\rm C\mkern-8mu\vrule height1.45ex depth-.05ex
width.05em\mkern9mu\kern-.05em}
{\rm C\mkern-8mu\vrule height1ex depth-.07ex
width.035em\mkern9mu\kern-.035em}
{\rm C\mkern-8mu\vrule height.65ex depth-.1ex
width.025em\mkern8mu\kern-.025em}}}

\def\RR{{\rm I\kern-1.6pt {\rm R}}}

\def\ZZ{{\rm Z}\kern-3.8pt {\rm Z} \kern2pt}
\def\IB{\relax{\rm I\kern-.18em B}}
\def\ID{\relax{\rm I\kern-.18em D}}
\def\II{\relax{\rm I\kern-.18em I}}
\def\IP{\relax{\rm I\kern-.18em P}}

\newcommand{\bear}{\begin{eqnarray}}
\newcommand{\eear}{\end{eqnarray}}

\def\6{\partial}

\def\bea{\begin{eqnarray}}
\def\eea{\end{eqnarray}}

\def\beqx{\begin{displaymath}}
\def\eeqx{\end{displaymath}}

\newcommand{\bmat}{\left(\begin{array}}
\newcommand{\emat}{\end{array}\right)}

\def\p{\pi}

\def\bo{{\raise-.3ex\hbox{\large$\Box$}}}
                                        
\def\pa{\partial}

\def\face{{\raise.2ex\hbox{$\displaystyle \bigodot$}\mskip-2.2mu \llap {$\ddot
        \smile$}}}

\def\>{\rangle}
\def\<{\langle}

\def\leftrightarrowfill{$\mathsurround=0pt \mathord\leftarrow \mkern-6mu
        \cleaders\hbox{$\mkern-2mu \mathord- \mkern-2mu$}\hfill
        \mkern-6mu \mathord\rightarrow$}
\def\dvec#1{\vbox{\ialign{##\crcr
        \leftrightarrowfill\crcr\noalign{\kern-1pt\nointerlineskip}
        $\hfil\displaystyle{#1}\hfil$\crcr}}}

\def\Tr{{\rm Tr \,}}

\def\-{\hphantom{-}}

\title{Dilaton EFT Framework For Lattice Data}

\author[a]{Thomas Appelquist}

\affiliation[a]{Department of Physics, Sloane Laboratory, Yale University, New Haven, Connecticut 06520, USA}

\author[a]{James Ingoldby}

\author[b]{Maurizio Piai}

\affiliation[b]{Department of Physics, College of Science, Swansea University, Singleton Park, SA2 8PP, Swansea, Wales, UK}

\date{\today}

\abstract{
	We develop an effective-field-theory (EFT) framework to analyze the spectra emerging from lattice simulations of a large class of confining gauge theories. Simulations of these theories, for which the light-fermion count is not far below the critical value for transition to infrared conformal behavior, have indicated the presence of a remarkably light singlet scalar particle. We incorporate this particle by including a scalar field in the EFT along with the Nambu-Goldstone bosons (NGB's), and discuss the application of this EFT to lattice data. We highlight the feature that data on the NGB's alone can tightly restrict the form of the scalar interactions. As an example, we apply the framework to lattice data for an SU(3) gauge theory with eight fermion flavors, concluding that the EFT can describe the data well.}

\begin{document}
\maketitle
\flushbottom

\section{Introduction}
\label{Sec:intro}

Lattice simulations of strongly interacting gauge theories indicate that infrared conformal behavior sets in
with a sufficiently large number of light fermions~\cite{latticeCFT,latticeCFT2,latticeCFT3,latticeCFT4,latticeCFT5,latticeCFT6,latticeCONF1,latticeCONF2,latticeCONF3,latticeUNC1,latticeUNC2,latticeUNC3,latticeUNC4,latticeUNC5} . In addition, a remarkably light singlet scalar
particle appears in the spectrum of recent simulations as this number is increased toward the critical value for
the transition to conformal behavior (the ``bottom of the conformal window") ~\cite{LSDresult,latticeSCALAR1,latticeSCALAR2,latticeSCALAR3,latticeSCALAR4}.
This has led to the suggestion that the light scalar should be interpreted as a dilaton, and that this interpretation could become even more accurate as the fermion number is taken closer still to the transition value.

Lattice simulations for gauge theories of this type have been carried out for a set of small fermion masses $m$, with extrapolation to $m = 0$ typically discussed
by fitting the results to continuum chiral perturbation
theory. This is equivalently an interpretation in terms of a chiral-Lagrangian
EFT consisting of (pseudo) Nambu-Goldstone bosons (NGB's) with a small mass $m_{\pi}^2 \propto m$.

A more general approach is to employ an EFT consisting of the NGB's together with a description of a light singlet scalar consistent with
its interpretation as a dilaton. Several authors have begun
this program~\cite{EFTDilaton1,EFTDilaton3,EFTDilaton4,EFTDilaton5,BLL}. In this paper, we develop such a framework for comparison with
existing lattice results as well as future simulations. Lattice results have so far been obtained for $m$ values such that the NGB mass is of the same order as the
scalar mass \cite{LSDresult,latticeSCALAR1,latticeSCALAR2,latticeSCALAR3,latticeSCALAR4}. These, in turn, are relatively small compared to the masses of other composite states, so that
the use of an EFT consisting of only these degrees of freedom should provide a good first approximation.
If and when simulations can be done at even smaller values of $m$, such that the NGB mass drops clearly
below the scalar mass, which in turn remains well below the other physical scales, the framework will
remain reliable.

The EFT we employ involves decay constants $f_d$ for the scalar and $f_{\pi}$ for the NGB's. In the EFT, $f_d$ enters as the order parameter for scale symmetry breaking. Both constants descend from the underlying, confining gauge theory with $m = 0$, and we expect them to have values set by the confinement scale.
A small scalar mass parameter $m_d$ also descends from the underlying, $m=0$ theory. Under proper conditions, to be discussed, quantum loop corrections are small, and are neglected in this paper. We instead provide a fit to existing lattice results, and a framework for future
simulations, employing the EFT at only the classical level. A notable feature of the framework is that lattice data on only the NGB's (their mass and decay constant), which are currently measured most precisely, are sufficient to determine a key parameter of the EFT and tightly restrict the form of the scalar potential.

In Section~\ref{Sec:EFT}, we describe the EFT, employing a general form for the scalar potential.
In Section~\ref {Sec:Comparison to Lattice Data}, we discuss the application of the EFT to lattice data and then describe a fit to current data from the LSD collaboration, drawing some conclusions about the
form of the EFT for that case. In Section \ref{Sec:The Small Mass-Shift Approximation}, we develop simple, linearized expressions to fit future lattice data over a small range of fermion masses $m$.
We summarize and conclude in Section~\ref{Sec:Summary and Conclusion}.

%%%%%%%%%%%%%%%%%%%%%%%%%%%%%%%%%%%%%%%%
\section{The EFT}
\label{Sec:EFT}

The low-energy EFT is built from the scalar field $\chi$ and
a set of NGB fields $\pi^a$. The latter arise from the spontaneous breaking of chiral symmetry, and the former, to the extent that
it can be interpreted as a dilaton,  arises from the spontaneous breaking of conformal symmetry.
The purely scalar part of the EFT consists of a kinetic term along with a
potential arising from the explicit breaking of conformal symmetry in the underlying theory, which we take to be
small:
\beqs
{\cal L}_d&=&\frac{1}{2}\partial_{\mu} \chi \partial^{\mu} \chi\,-\,V(\chi)\,.
\label{Eq:Ld}
\eeqs
We assume that the potential has a minimum at some value $f_{d}$,
and that it is comparatively shallow, so that the mass $m_d$ of the fluctuations around the minimum satisfies $m_d\ll 4\pi f_{d}$.

A specific choice of the potential amounts to supplementing the EFT with partial information from the underlying dynamics.
Two examples from the literature are
\beqs
V_1&=&\frac{m_d^2}{2f_{d}^2}\left(\frac{}{}\frac{\chi^2}{2}-\frac{f_{d}^2}{2}\right)^2\, \label{v1},\\
V_2&=&\frac{m_d^2}{16 f_{d}^2}\chi^4\left(4\ln\frac{\chi}{f_{d}}-1\right)\, \label{v2},
\eeqs
normalized such that in each case $m_d$ is the scalar mass.
The first is a weakly-coupled potential such as the one appearing in the standard model. It can arise from the deformation of an underlying conformal field theory (CFT)  by relevant operators. 
The second has been proposed in Ref. \cite{GGS} as a way to model the behavior of a CFT deformed  by a
nearly marginal operator (see also Ref. \cite{EFTDilaton1}). Unlike in previous approaches \cite{EFTDilaton1,EFTDilaton3,EFTDilaton4}, we do not make an assumption about the specific functional form of the potential and instead allow it to be determined by the lattice data. This is a key element of novelty in our framework.

The NGB's arising from the spontaneous breaking of chiral symmetry are described in terms of
a field $\Sigma$ transforming as $\Sigma\rightarrow U_L \Sigma U_R^{\dagger}$, with $U_L$ and $U_R$ the matrices of $SU(N_f)_L$ and $SU(N_f)_R$ transformations. (This approach can be adapted to other symmetry groups and breaking patterns). The $\Sigma$ field satisfies the nonlinear constraint $\Sigma \Sigma^\dagger = \mathbb{I}$. We hence write:
\beqs
{\cal L}_{\pi}&=&\frac{f_{\pi}^2}{4}\left(\frac{\chi}{f_{d}}\right)^2 \,\Tr\left[\partial_\mu \Sigma (\partial^{\mu} \Sigma)^{\dagger}\right]\,,
\label{Eq:Lpi}
\eeqs
where the coupling to the dilaton field (introduced here as a compensator field to maintain the scale invariance of this term in the Lagrangian) is dictated by the fact that the $\Sigma$ kinetic term has scaling dimension $d=2$.
The $\Sigma$ field can be parametrized through
$\Sigma=\exp\left[2i \pi/f_{\pi} \right]$ where $\pi=\sum_a\pi^a T^a$ and $T^a$ are the $N_f^2 - 1$ generators of $SU(N_f)$ normalized
as $\Tr [ T^aT^b]=\frac{1}{2}\delta^{ab}$. In contrast with the linear-sigma-model description of chiral symmetry breaking, more generally $f_d$ and $f_\pi$ are independent, as the underlying strong dynamics may involve condensates besides the chiral-symmetry-breaking one.

In lattice calculations of particle masses and decay constants in the underlying gauge theory, chiral symmetry (as well as conformal symmetry) must be explicitly broken by the introduction of a small fermion mass term of the form $m\bar{\psi}\psi$. The explicit breaking is implemented in the EFT through the term

\beqs
{\cal L}_M&=&\frac{ m_{\pi}^{2} f_{\pi}^2 }{4}\left(\frac{\chi}{f_{d}}\right)^y\,\Tr \left[\Sigma + \Sigma^{\dagger}\right] \,,
\label{Eq:LM}
\eeqs
where $ m_{\pi}^{2}  = 2 m B_{\pi}$, with $B_{\pi}$ determined by the chiral condensate of the underlying theory ($B_\pi=\langle\bar{\psi}\psi\rangle / 2f_{\pi}^2$). The product $m B_{\pi}$ is RG-scale independent, with each factor typically defined at the UV cutoff (the lattice spacing). The parameter $y$ has been argued to be the scaling dimension of $\bar{\psi}\psi$ in the underlying theory \cite{BLL}. This scaling dimension is an RG-scale dependent quantity, which could vary from $3$ at UV scales where the theory is perturbative to smaller values near the confinement scale. Analyses of near-conformal theories have suggested a scaling dimension $ \approx 2$ at this scale \cite{Georgi}. We keep $y$ as a free parameter to be fitted to the lattice data.

 Expanding ${\cal L}_M$ around $\pi^a = 0$ gives
\beqs
{\cal L}_M &=& \frac{N_f m_{\pi}^2 f_{\pi}^2}{2} \left(\frac{\chi}{f_d}\right)^{y}  -
 \frac{m_{\pi}^2}{2}\left(\frac{\chi}{f_d}\right)^{y} \pi^{a}\pi^{a} \,+\,\cdots \,,
\eeqs
generating a negative contribution to the scalar potential as well as an NGB mass term.
The new contribution to the scalar potential shifts both the VEV and the mass of the scalar field $\chi$. The shifted VEV will, through Eq.~(\ref{Eq:Lpi}), re-scale the NGB kinetic term, and hence the NGB decay constant.

%%%%%%%%%%%%%%%%%%%%
\section{Comparison To Lattice Data}
\label{Sec:Comparison to Lattice Data}

\subsection{ General Discussion}

Lattice simulations are currently carried out for $SU(N_c)$ gauge theories with fairly small $N_c$ ($= 2,3$). For these cases, $N_f$
cannot be too large if the theory is to be in the confining phase.
Our program is to use the full EFT ${\cal L}_{d}+{\cal L}_{\pi}+{\cal L}_M$ to describe current and future lattice results for the NGB's and the light $0^{++}$ scalar, the latter having already been observed for example in the $N_f =8$ $SU(3)$ simulations. The parameters $f_d$, $f_{\pi}$, $y$, and the scalar potential $V(\chi)$, have no dependence on the fermion mass and are  held fixed as the parameter $m_{\pi}^2=2 mB_{\pi}$ is varied.
In this paper, this will be done using the EFT at only the classical level. We discuss this approximation further in Section~\ref{Sec:Summary and Conclusion}.

When comparing the predictions of our EFT to the lattice data, we assume throughout that lattice discretization and finite volume effects are small. We therefore add no additional terms to the EFT Lagrangian to represent such effects. Neglecting these lattice artifacts should introduce only a small systematic error.

A set of $m_{\pi}^2$-dependent quantities $F_d^2$, $M_d^2$,  $F_{\pi}^2$ and $M_{\pi}^2$ emerge from a tree level analysis of the EFT. The quantity $F_d$ is defined to be the $\chi$-value that minimizes the full potential 
\beqs
W(\chi) = V(\chi) - (N_f m_{\pi}^2 f_{\pi}^2 /2) (\chi /f_d)^{y}.
\label{Eq:Wdef}
\eeqs
$F_d$ is finite assuming only that $V(\chi)$ is stable and increases at large $\chi$ more rapidly than $\chi^y$. In the case of the potential $V_1$ given by Eq.~(\ref{v1}), $F_d$ is determined by the equation
\beqs
      \left(\frac{F_d^2}{f_d^2}\right)^{2-y/2}  \left[ 1-   \left(\frac{f_d^2}{F_d^2}\right) \right]= \frac{yN_f f_\pi^2}{f_d^2}\left(\frac{m_{\pi}^2}{m_{d}^2}\right) \, \label{decayconstv1},
\eeqs 
whereas for the potential $V_2$ in Eq.~(\ref{v2}), it is determined by 
\beqs
\left(\frac{F_d^2}{f_d^2}\right)^{2-y/2}\ln\left(\frac{F_d^2}{f_d^2}\right) = \frac{yN_f f_\pi^2}{f_d^2}\left(\frac{m_{\pi}^2}{m_{d}^2}\right).
\label{decayconstv2}
\eeqs

In general, $F_d $ depends on the interplay between the two parts of the potential $W(\chi)$. The physical scalar mass $M_d^2$ is determined by the curvature of the full potential at its minimum.
The remaining two quantities, $F_{\pi}^2$ and $M_{\pi}^2$, can be identified after properly normalizing the NGB kinetic term.
They are given in general by simple scaling formulae:
\beqs
 \frac{F_{\pi}^2}{f_{\pi}^2}&=&\frac{F_{d}^2}{f_{d}^2}\,,\label{Eq:fpi/fd}\\ \frac{M_{\pi}^2}{{m}_{\pi}^2}&=&\left(\frac{F_{d}^2}{f_{d}^2}\right)^{y/2-1} \label{Eq:Mpiscaling}
\eeqs
(see also Ref.~\cite{EFTDilaton1}). For a given value of $N_f$, the dependence  of each of the four quantities $F_{d}^2$, $M_d^2$, $F_{\pi}^2$, and $M_{\pi}^2$ on $m_{\pi}^2 \equiv 2 m B_{\pi}$  is described in terms of the four parameters $f_d$, $f_{\pi}$, $m_d^2$, $y$, and whatever additional parameters enter the scalar potential $V(\chi)$. One immediate prediction is that $F_{d}$ and $F_{\pi}$ have the same functional dependence on $m_{\pi}^2$.

We stress that $F^2_\pi/f^2_\pi$ and the other ratios in the scaling relations (\ref{Eq:fpi/fd}) and (\ref{Eq:Mpiscaling}) are not restricted to be close to unity. These ratios, entering at the classical level, can become large due to the increase of $F_\pi$ and $F_d$ with $m^2_\pi/m^2_d$ (see Eqs. (\ref{decayconstv1}) and (\ref{decayconstv2})). The ratio $M_d^2/ m_d^2$ also increases in this limit.  Importantly though, quantum loop corrections can be small even when these ratios are large. The quantum corrections depend on the quantities $M^2_\pi/(4\pi F_\pi)^2$ and $M^2_d/(4\pi F_d)^2$. These can remain small when each of the capitalized scales increases as $m^2_\pi /m_d^2$ is increased. The upper limit on the range of validity of the EFT, determined by $4\pi F_\pi$ and $4\pi F_d$, increases commensurately.

The four capitalized quantities are directly related at tree level to physical processes involving the NGB's and the scalar. Three of them are measured by lattice studies of the underlying, microscopic gauge theory.
The masses $M_{\pi}$ and $M_{d}$ can be found by measuring the exponential fall of appropriate correlation functions, and $F_{\pi}$ can be extracted from simulations of the axial-vector current correlator. It is defined using the same conventions as in \cite{LSDresult}. The extraction of $F_d$ from a lattice measurement of a correlation function in the underlying gauge theory has not yet been reported. The connection between correlation functions in the gauge theory and the $F_d$ of our EFT requires further renormalization analysis. While $F_d$ enters our framework as the VEV of the scalar field, we do not require its numerical value in our fit to the LSD data.

The comparison to lattice data will focus first on the quantities  $F_{\pi}^2$  and  $M_{\pi}^2$, which are currently known most precisely. For this purpose, it is helpful to note that
the two scaling relations, Eqs.~(\ref{Eq:fpi/fd}) and (\ref{Eq:Mpiscaling}) give
\beqs
M_{\pi}^2 (F_{\pi}^2)^{(1-y/2)} = C m \,,
\label{yscaling}
\eeqs
where $C = 2 B_{\pi}  (f_{\pi}^2)^{(1-y/2)}$. Fitting lattice data to Eq.~(\ref{yscaling}) can allow an accurate determination of $y$.

Another key question is to what extent the form of the scalar potential $V(\chi)$ can be
determined by a fit to lattice data.  With the small amount of data available so far, only limited progress can be made on this ``inverse-scattering'' problem. 
We will find it helpful, even with the current data,  to 
consider the slope of the scalar potential $V(\chi)$ at the value of $\chi$  ($\chi=F_d$) that minimizes the full potential $W(\chi)$.  From Eqs.~(\ref{Eq:Wdef}) and (\ref{Eq:fpi/fd}),
\beqs
\left.\frac{\pa V}{ \pa \chi}\right|_{\chi=F_d} = \frac{yN_{f}f_{\pi}^2}{2f{_d^2}} M_{\pi}^{2} F_{d} = \frac{yN_{f}f_\pi}{2f{_d}} M_{\pi}^{2} F_{\pi}\, .
\label{dV}
\eeqs
Since $F_{\pi} \propto F_d$, a plot of the data for $ M_{\pi}^{2} F_{\pi}$ versus $F_{\pi}$  provides a measure of the slope of $V(\chi)$  at $\chi = F_d$ versus $F_d$ itself. This slope vanishes in the chiral limit $m=0$, corresponding to $F_{\pi} = f_{\pi}$, since then $F_d = f_d$ (the minimal point of $V(\chi)$ itself). 
 As $F_{\pi} (\propto F_d)$  is increased,
the slope of $V(F_d)$ increases through positive values. We use Eq.~(\ref{dV}) to analyze data from the LSD collaboration in the next sub-section.

This procedure can be taken to the next stage by bringing the lattice data on $M_d^2$ into the analysis. From Eqs.~(\ref{Eq:fpi/fd}), (\ref{Eq:Mpiscaling}) and (\ref{dV}), together with the definition $M_d^2 \equiv \pa^2W/\pa \chi^2|_{\chi=F_d}$, one can derive an expression for the second derivative of $V$ at $\chi = F_d$:
\beqs
\left.\frac{\pa^2 V}{\pa \chi^2}\right|_{\chi=F_d} = \quad M^2_d + \frac{y(y-1)N_f f^2_\pi m^2_\pi}{2f^2_d}\left(\frac{F^2_d}{f^2_d}\right)^{y/2-1} = \quad M^2_d + \frac{y(y-1)N_f f^2_\pi}{2f^2_d}M^2_\pi.
\label{d2V}
\eeqs
Thus data for $M_d^2$ could be used in the analysis alongside the $M^2_\pi$ and $F^2_\pi$ data, to allow a fit that can better constrain both the scalar potential, and the other free parameters of the Lagrangian.

\subsection{Application to the LSD Data}

\begin{figure}[h]
	\centering 
	\subfigure[\hspace{0.2cm}The squared masses of the NGB's and scalar.]{\includegraphics[width= 7cm]{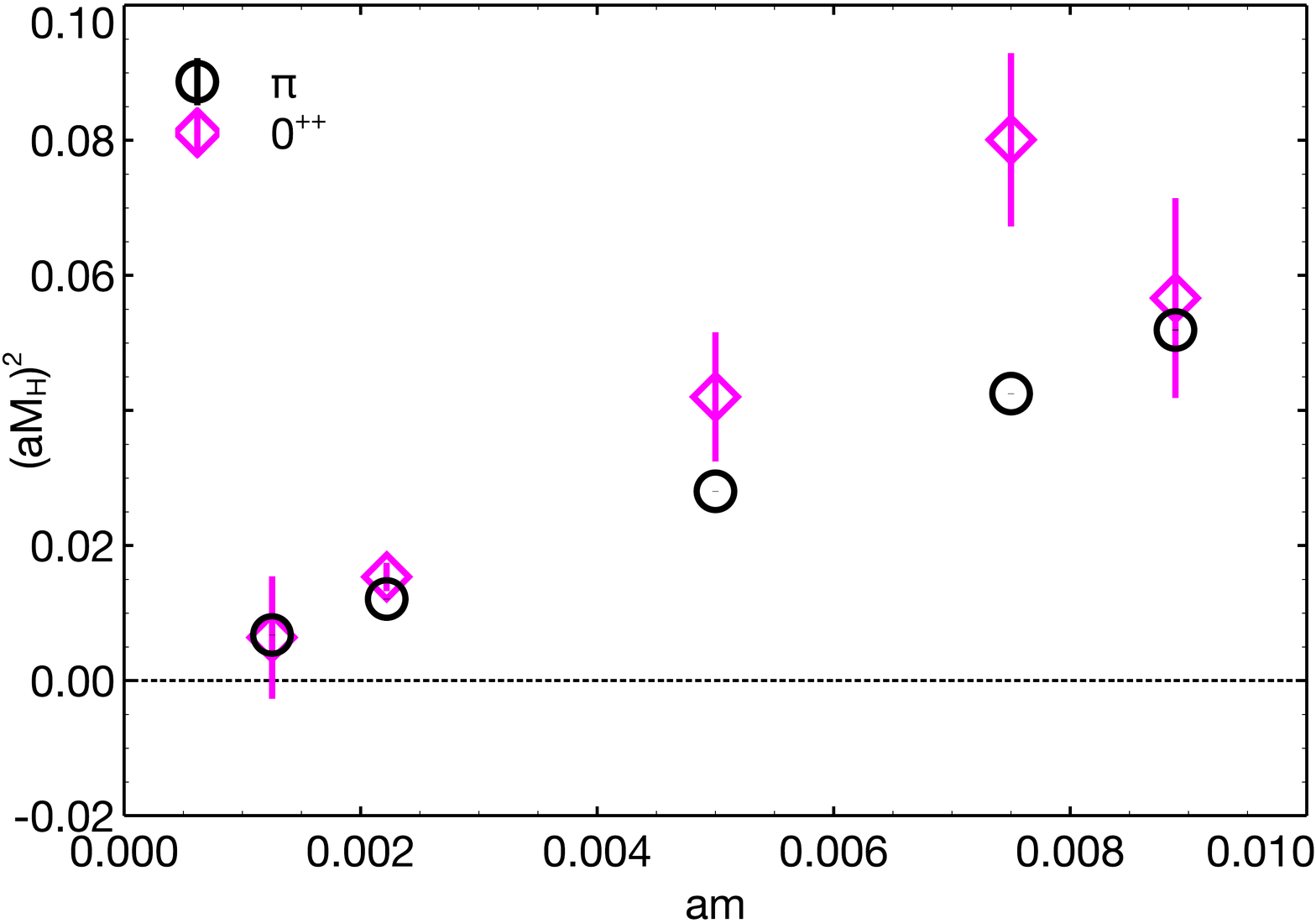}} \qquad \subfigure[\hspace{0.2 cm}The squared NGB decay constant.]{\includegraphics[width=7cm]{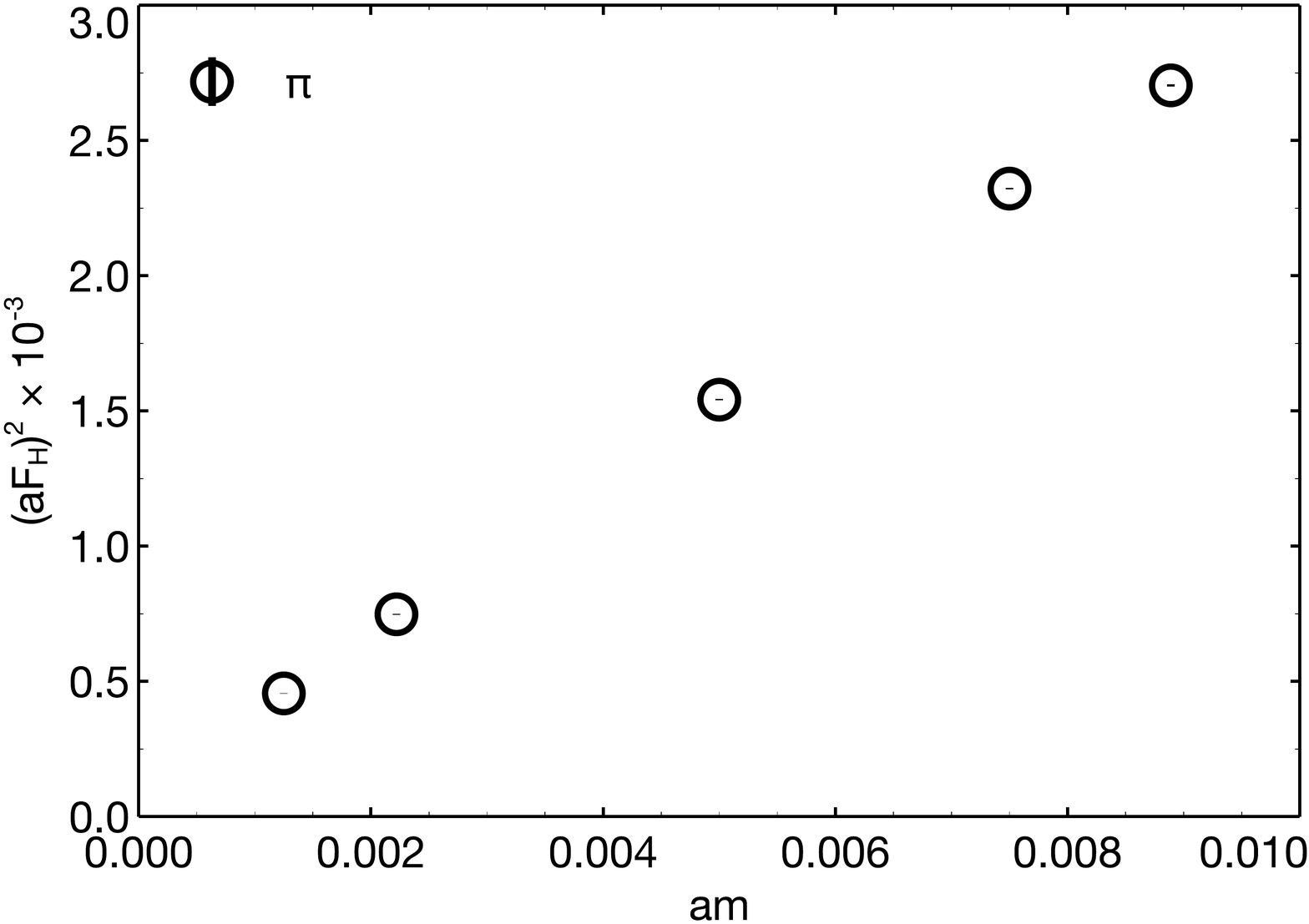}}  
	\caption{Data from the LSD collaboration \cite{LSDresult,LSDlatest}. Error bars represent only the statistical uncertainty in the data. The lattice spacing is denoted by the parameter ``a".} 
	\label{2figs}
\end{figure}

We next apply our EFT framework to the LSD collaboration data  for the $SU(3)$ gauge theory with $N_f = 8$ \cite{LSDresult}. These data, which cover the smallest fermion mass range studied as of yet for this theory, are currently limited to $M_d^2$, $F_{\pi}$, and $M_{\pi}^2$. They are shown in 
Figs. 1a and 1b. A list of the numerical values and errors has been provided to us by the LSD collaboration. We first note that the lattice data for $M_{\pi}^2$ and $F_{\pi}^2$ are remarkably linear throughout the range of $m$ values. We also note that $M^2_\pi/(4\pi F_\pi)^2\ll1$ throughout the range of the data, indicating that loop corrections are small. The data for $M_d^2$ are compatible with linearity but the errors are large. The  $M_{\pi}^2$  data are consistent with an expected intercept of $0$. A finite intercept is expected in the case of the $F_{\pi}^2$ data.

The linearity of the $M_{\pi}^2$ data combined with the substantial variation of 
$F_{\pi}^2$  with $m$ leads, through the scaling relation Eq.~(\ref{yscaling}), to a determination of $y$ and $C$. Since the data for $M_{\pi}^2$ is itself near-linear in $m$ and $F_{\pi}^2$  is varying substantially, $y$ must be close to $2$. We fit the data assuming that an additional, conservative  $2\%$ systematic uncertainty should be assigned to it, for both  $F_{\pi}^2$ and $M_{\pi}^2$. This is consistent with the estimate of finite-volume and lattice-discretization artifacts reported in Ref. \cite {LSDresult}. In addition, there are systematic uncertainties associated with the EFT we employ. We discuss these briefly in Section \ref{Sec:Summary and Conclusion}, but do not include them in our fit. The result of our fit to Eq.~(\ref{yscaling}), treating both $y$ and $C$ as free parameters, is 
 \beqs
 y = 2.1 \pm 0.1 ,
 \label{yresult}
 \eeqs
 with $1\sigma$ uncertainty and $\chi^2/ N = 0.34$ (where $N=3$). The fit value for $C$ is $7.2\pm1.8$. The result for $y$ is not inconsistent with $y=2$  and therefore with $M_{\pi}^2 = m_{\pi}^2 \equiv 2 B_{\pi} m$ (the zeroth-order chiral perturbation theory formula for $M^2_\pi$)\footnote{If the lattice data for $M_{\pi}^2$ were not so linear in $m$, they could still be consistent with the scaling relation Eq.~(\ref{yscaling}), but with $y\not\approx 2$.}. By contrast, the substantial variation of $F_{\pi}$ with $m$ looks nothing like 
 zeroth-order chiral perturbation theory. In our EFT, its variation with $m$ is naturally 
 accommodated at the classical level.

\begin{figure}
	\centering
	\includegraphics[width=10cm]{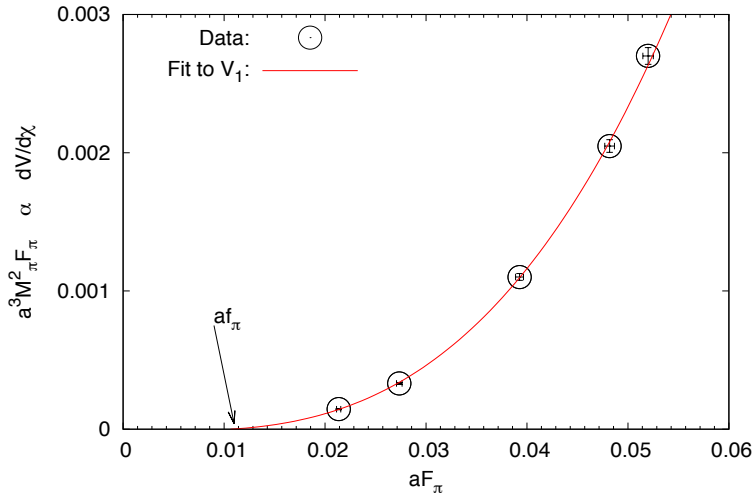}
	\caption{Lattice data for the product $M^2_\pi F_\p$ (proportional to the slope of the scalar potential at $\chi = F_d$)  versus $F_{\pi}$ (proportional to $F_d$.)  The red line represents a fit to Eq.~(\ref{v1fit}) which derives from the potential $V_1$. The lattice spacing is denoted by the parameter ``a".}
	\label{fig2}
\end{figure}

The near-linearity with $m$ of the $F_{\pi}^2$ data  provides more detailed information. Through Eq.~(\ref{Eq:fpi/fd}), it implies that $F_d^2$ must also be near linear in $m$. This suggests a relation similar to that of Eq.~(\ref{decayconstv1}) which arises from the  $V_1$ potential, together with $y \approx 2$, but it doesn't rule out other forms for the potential. To proceed, we use Eq.~(\ref{dV}) relating the slope of $V(\chi)$ at $\chi = F_d$ to the product $M_{\pi}^{2} F_{\pi}$.   In Fig.~\ref{fig2}, we plot the LSD data for $M_{\pi}^{2} F_{\pi}$ against $F_{\pi}$ . Error bars representing the 2\% systematic uncertainty are shown. Since each point on the vertical axis is proportional to the slope of $V(\chi)$ at $\chi = F_d$ and each point on the horizontal axis is proportional to $F_{d}$, the points in the figure display the shape of the scalar potential $V(\chi)$ for the $N_f = 8$ theory. 

The data indicate clearly that $V(\chi)$ increases with $\chi$ for a range of $\chi$ beyond its minimum at a rate much faster than $\chi^2$, confirming that the scalar sector of the EFT is self-interacting.  The data are in fact consistent with the large-$\chi$ behavior $V(\chi) ~\propto \chi^4$ as in $V_1$ Eq.~(\ref{v1}). For this potential, Eqs.~(\ref{Eq:fpi/fd}) and (\ref{dV}) give

\beqs
M^2_\pi F_\pi  = \frac{F_\pi}{A}(F_\pi^2 - f^2_\pi),
\label{v1fit}
\eeqs
where $A\equiv(yN_f f^4_\pi)/(m^2_d f^2_d)$. The data can be fit to this form, with $f_\pi$ and $A$ treated as independent parameters. The best fit is represented by the red line in Fig.~\ref{fig2}. The fit parameters are $af_{\pi} = 0.01\pm0.002,\,\,\, A =0.05\pm0.005,\,\,\, \chi^{2}/N=1.1 $ (where $N=3$). From Fig.~\ref{fig2}, it can also be seen that $F^2_\pi/f^2_\pi\gg1$ throughout the range of the data.

A fit deriving from other forms of the scalar potential qualitatively similar to $V_1$ is also possible. An example is $V_2$, for which 
\beqs
M^2_\pi F_\pi  = \frac{F_\pi^3}{A}  \ln \left(\frac{F_\pi^2}{f^2_\pi}\right). \,
\eeqs
This can also lead to a good fit, but only with a smaller value of $f_{\pi}$. Here, we don't show this fit or others based on alternative forms of the potential. While potentials qualitatively unlike $V_1$ can be ruled out, the limited amount of data available does not yet allow us to distinguish between a variety of similar forms.  As more data points become available, spread over a larger range of fermion masses, the above method can be used to determine the functional form of the scalar potential with increasing precision, over a larger range of field values. We note again that the NGB lattice data alone ($F_{\pi}^2$  and $M_{\pi}^2$) can provide this information.

To take this analysis further, the lattice data for $M_d^2$ shown in Fig. 1a can also be included in the fits. Because of the large statistical errors currently associated with these points, they don't yet add precision to the analysis of the form of the scalar potential, but they are sufficient to provide some 
approximate information about the parameter $f_d$ and the associated physical quantity $F_d$. It can be seen that for $y \sim 2$ and for any potential with the large-$\chi$ behavior of $V_1$ or $V_2$, the relation $M_d^2 \sim N_{f} f_{\pi}^{2} m_{\pi}^{2} / f_{d}^2$ holds in the range of the lattice data. Using the fact that $m_\pi^2 \approx M_\pi^2 \sim M_d^2 $ in this range, we have the rough prediction $F_d^2 / F_{\pi}^2  = f_d^2 / f_{\pi}^2 \sim N_f$ throughout the range. In the future, more information about the potential $V$ can also be found by including $M_d^2$ data. Eq.~(\ref{d2V}) then provides a measure of the second derivative of $V$ at the minimum $F_d$ of the full potential $W$.

It is important to note that while the parameter $y$ can be accurately determined directly from lattice data using Eq.~(\ref{yscaling}), the parameters $f_{\pi}$, $f_{d}$, and $m_{d}$ are extrapolated quantities. The size of these parameters depends upon the form of the scalar potential in the vicinity of its minimum. The increasingly accurate determination of $V(\chi)$ will require lattice data at decreasingly small values of $m$. The current lattice data lie in a regime where $F_\pi/f_\pi \gg 1$, $F_d/f_d \gg 1$ and $M_d^2/m_d^2 \gg1$. The second term in the potential $W(\chi)$  (Eq.~\ref{Eq:Wdef}) begins to dominate the mass term in $V(\chi)$ since $m^2_\pi\approx M^2_\pi \gg m^2_d$.

%%%%%%%%%%%%%%%%%%%%
\section{Small Mass-Shift Approximation - A Side Note}
\label{Sec:The Small Mass-Shift Approximation}

Looking to the future, lattice data for the $SU(3)$ gauge theory with $N_f = 8$ could extend to smaller $m$ values as well as include more densely spaced points in the range of Figs. 1a and 1b. There will also be data for $F_d^2$ as a function of $m$. Simulations of other theories could  produce additional interesting data for each of the masses and decay constants. These results could appear linear as a function of $m$ or exhibit  nonlinear behavior. For future analysis of such data sets using our EFT and allowing for a general form of the scalar potential $V(\chi)$, it could be helpful to linearize the physical quantities about a reference value $ m_r \equiv m_{\pi\,r}^2 /2  B_{\pi}$. In this section, we briefly describe this approach.

With $m$ restricted to a small enough neighborhood of $m_r$,  the quantities of interest will be sensitive to the shape of the full potential $W(\chi)$ only in the neighborhood of its minimum with $m = m_r$. The full potential can therefore be approximated as
\beqs
W(\chi) & = W_{r}(\chi) - \frac{N_f \Delta m^2_\pi f^2_\pi}{2}\left(\frac{\chi}{f_d}\right)^y,
\eeqs
with
\beqs
W_{r}(\chi) & \approx \frac{1}{2}M^2_{d\,r}(\chi -F_{d\,r})^2 + \frac{g_r}{3!}\frac{M^2_{d\,r}}{F_{d\,r}}(\chi - F_{d\,r})^3 + \dots \, ,
\eeqs
where $F_{d\,r}$ is the minimum of the scalar potential for the reference value $m_r$ and 
$\Delta m_{\pi\,r}^2 \equiv 2 \Delta m B_{\pi} \equiv 2 (m -m_{r} )B_{\pi}$. $M_{d\,r}^2$ is the scalar mass at the reference value  and $g_r$ is a free parameter controlling the strength of the scalar cubic self-interaction. We expect it to be $ O(1)$.

We make the replacements

\begin{align}\nonumber
f_{\pi}^2 = F_{\p\,r}^2 \frac{f_{d}^2}{F_{d\,r}^2}, \qquad m^{2}_{\pi\,r} = M^2_{\pi\,r}\left( \frac{f_{d}^2}{F_{d\,r}^2}\right)^{y/2-1}, \qquad
\Delta m^{2}_\pi =  M^2_{\pi\,r}  \frac{\Delta m}{m_r} \left(\frac{f_{d}^2}{F_{d\,r}^2}\right)^{y/2-1}\,,
\end{align}
where $\Delta m = m - m_r$.   The quantities $F_{d}^2$, $M_d^2$, $F_{\pi}^2$, and $M_{\pi}^2$  then have the following dependence on $\Delta m / m_r$:

\begin{align}
\frac{F_d^2}{F_{d\,r}^2} & = 1 + 2 \alpha_r \frac{ \Delta m}{m_{r}}  +   O\left(\Delta m^{2}\right),\\
\frac{M^2_d}{M^2_{d\,r}} & = 1 + \alpha_r  (g_r + 1 - y)\frac{\Delta m}{m_{r}} + O\left(\Delta m^2\right)\\
\frac{F_{\pi}^2}{F_{\pi\,r}^2} &= \frac{F_d^2}{F_{d\,r}^2},\\
\frac{M_{\pi}^2}{M_{\p\,r}^2} & = 1+ \left[ 1 + \alpha_{r}(y-2)  \right] \frac{\Delta m}{m_r} +   O\left(\Delta m^2\right)\,,
\end{align}
where 
\beqs
\alpha_r = \frac{ y N_{f} F^2_{\pi\,r} M^2_{\pi\,r}} { 2 F^2_{d\,r}M^2_{d\,r}}.
\eeqs

One can fit lattice data as a function of $\Delta m / m_r$ using these formulae and their extensions to higher order. The expansion is reliable providing $\alpha_{r} \Delta m / m_r \ll1$. 
The four parameters $F_{d\,r}^2$, $M_{d\,r}^2$, $F_{\pi\,r}^2$, and $M_{\pi\,r}^2$ are simply the values of $F_{d}^2$, $M_{d}^2$, $F_{\pi}^2$, and $M_{\pi}^2$ at the reference point. The additional two  parameters $y$ and $g_r$ can be determined by the slope of the curves at the reference point. At  higher orders, additional parameters describing the shape of the potential $W(\chi)$ will enter. 

The parameter $g_r$ is itself sensitive to the shape of the potential. In the absence of the chiral-symmetry-breaking second term in Eq.~(\ref{Eq:Wdef}),  we have $g_r = 3$ for $V = V_1$ and $g_r= 5$ for $V = V_2$. Away from the chiral limit, at some reference value $m_r$, the contribution of the second term must be taken into account. The value of $g_r$ will depend on the shape of $V$, the value of $y$ and the other parameters, and the choice of the reference mass $m_r$.  For the case $ V =V_1$, it will remain the case that $g_r =3$ if $y=2$.

%%%%%%%%%%%%%%%%%%%%
\section{Summary and Conclusion}
\label{Sec:Summary and Conclusion}

We have developed a simple EFT framework for the interpretation of lattice results for confining gauge theories, in which the light-fermion count is near to but below the critical value for transition to conformal behavior. The lattice studies indicate that a remarkably light scalar appears in the spectrum along with the NGB's and higher-mass states. Interpreting the scalar as a dilaton, we have included only it and the NGB's in the EFT, and allowed a general form for the dilaton potential. 

The presence of a small fermion mass $m$  in the underlying gauge theory, necessary for lattice simulations, leads to  a chiral-symmetry-breaking term in the EFT. The coupling of this term to the scalar field 
is described by a parameter $y$, to be fit to lattice data. We provided expressions for the masses and decay constants of the scalar particle and NGB's  appropriate for comparison to lattice data, noting that the data can be used to determine $y$ as well as the shape of the scalar potential above its minimum. 

We applied this framework at the classical level to the current LSD collaboration data for an $SU(3)$ gauge theory with $N_f=8$, which covers the smallest fermion-mass range studied for this theory. Even with the limited data available so far, we concluded generally from a fit to the data for $F_{\pi}^2$, and $M_{\pi}^2$ that $y \approx 2$ and that the scalar potential $V(\chi)$ grows approximately like $\chi^4$ beyond its stable minimum. Among the other parameters $f_d$, $f_{\pi}$, $m_d$ of our EFT, we have so far provided only an estimate of $f_{\pi}$ (following Eq.~(\ref{v1fit})), for the case of the $V_1$ potential . The data for $M_d^2$ are currently less accurate than for $F_{\pi}^2$, and $M_{\pi}^2$. We have used them so far only to predict roughly that  $F_d/ F_{\pi} = f_d/ f_{\pi} \sim \sqrt{N_f}$. This is consistent with our starting assumption that the scalar particle is weakly self-interacting, that is $m_d \ll 4 \pi f_d$. As more data points become available, our method can be used to determine the functional form of the scalar potential with increasing precision, over a larger range of field values, and to extract more accurately the chiral-limit parameters $f_d$, $f_{\pi}$, $m_d$.

For purposes of analyzing future lattice data, we developed expressions for the masses and decay constants of the scalar and NGB's, linearized in $m$ about a reference value $m_r$. This framework is well suited to analyze future data that are dense in the neighborhood of a reference value. The scalar potential is obtained from data as a Taylor series, making it possible to exclude potentials that are inconsistent with data in a systematic way.

The EFT we have employed neglects the effects of heavier states such as the vector and axial-vector  bound states  produced by the underlying gauge theory. In the case of the LSD data  for the $SU(3)$ gauge theory with $N_f=8$, the masses of these states have been measured for each of the $m$ values in 
Fig. 1. Throughout this range, $M_{\pi}^2 / M_{V}^2 \leq 0.2 $ dropping to $\leq 0.1$ for the lowest value. The data for $M_d^2$, with their larger statistical errors,  also satisfy a similar bound. The axial state is still heavier leading to corrections that are even more suppressed. 

Our framework has also neglected higher order corrections in perturbation theory arising from loops of NGB's and the scalar. By inspection of the lattice data in Fig. 1, one can see that for all but one of the points, $M^2_\pi \sim M^2_d$. It is also the case for all of our fits that throughout the range, $ F^2_\pi \lesssim F^2_d$. Thus, from the data, one can see that the loop-expansion quantities $M^2_\pi/(4\pi F_\pi)^2$ and $M^2_d/(4\pi F_d)^2$ are small. The loop expansion also has counting factors that can grow with $N_f$, as well as chiral logarithms, and these have to be included in a full analysis of these corrections. This is beyond the scope of the present paper. We note here only that the order of magnitude of these corrections varies very little throughout the mass range of Figs. 1a and 1b, so that their systematic effect should be possible to control. Loop level effects, and the effects of heavier states can be incorporated into higher-dimension operators correcting our EFT that are suppressed by a cutoff scale $\Lambda \sim M_V$.

More generally, our EFT framework can  be applied to lattice data from any strongly coupled gauge theory with a light-fermion count below the bottom of the conformal window, but close enough to exhibit a light scalar in the spectrum. A current example could be the $ SU(3)$ gauge theory with a doublet of fermions in the symmetric-tensor representation. Our framework and analysis can be refined further as the amount and quality of lattice data increases, with the ultimate goal of a full ``inverse-scattering" reconstruction of the scalar potential from the data.

%%%%%%%%%%%%%%%%%%%%%%%%%%%%%%%%%%%%%%%
\vspace{1.0cm}
\begin{acknowledgments}
We thank Enrico Rinaldi for his assistance in providing the latest data and plots from the LSD collaboration. We also thank Biagio Lucini, Pavlos Vranas, George Fleming and Andrew Gasbarro for helpful discussions. The work of TA and JI is supported by the U.S. Department of Energy under the contract DE-FG02-92ER-40704. The work of MP is supported in part by the STFC Consolidated Grant ST/L000369/1.
\end{acknowledgments}
\vspace{1.0cm}

%%%%%%%%%%%%%%%%%%%%%%%%%%%%%%%%%%%%%%%%

\end{document}